\documentclass{elsart}

\input epsf
\input psfig

\def\xslide#1#2#3#4#5#6{\centerline{\psfig
{figure=#1,height=#2,bbllx=#3bp,bblly=#4bp,bburx=#5bp,bbury=#6bp,clip=}}}

\begin{document}

\begin{frontmatter}

\hfill Preprint INP 1803/PH

\vspace{0mm}

\title{$\omega \to \pi \pi$ decay in nuclear medium%
\thanksref{grants}}
\thanks[grants]{Research supported by PRAXIS grants XXI/BCC/429/94 and
        PCEX/P/FIS/13/96, and by
        the Polish State Committee for
        Scientific Research grant 2P03B-080-12.}
\thanks[emails]{E-mail: broniows@solaris.ifj.edu.pl,
 florkows@solaris.ifj.edu.pl, brigitte@fteor5.fis.us.pt}

\author[INP]{Wojciech Broniowski},
\author[INP]{Wojciech Florkowski}, and
\author[Coimbra]{Brigitte Hiller}

\address[INP]{H. Niewodnicza\'nski Institute of Nuclear Physics,
         PL-31342 Krak\'ow, Poland}

\address[Coimbra]{Center for Theoretical Physics,
   University of Coimbra, P-3000 Coimbra, Portugal}

\begin{abstract}
We calculate the width for the $\omega \to \pi\pi$ decay in nuclear medium.
Chiral dynamics and low-density approximation are used. At densities
around twice the nuclear saturation density we estimate the
partial width for the decay of the
longitudinal mode to be of the order of a few tens
of MeV, and for the transverse mode a few times less.
\end{abstract}

\begin{keyword}
relativistic heavy-ion collisions, vector mesons, in-medium
modifications of particle properties
\end{keyword}

\end{frontmatter}
\vspace{-7mm}

PACS: 25.75.Dw, 21.65.+f, 14.40.-n


Recent relativistic heavy-ion experiments have brought evidence that nuclear
medium modifies substantially properties of meson excitations. In
particular, the dilepton measurements in the CERES \cite{ceres} and HELIOS 
\cite{helios} experiments indicate that either the positions of light vector
mesons are shifted down or their widths are increased. Such a behavior is
expected from many theoretical calculations \cite
{celenza,jean,cassing,li,hatsuda,klingl,friman1,eletsky,friman2} (for review
see \cite{hadrons,heidelberg}). Among the sources of the in-medium
modifications of meson properties are phenomena forbidden in the vacuum,
which become possible in the presence of nuclear medium. In particular,
constraints such as $G$-parity are no longer effective for mesons moving
with respect to nucleons in nuclear matter.

In this Letter we analyze an example of such an ``exotic'' process: the
in-medium decay of $\omega \to \pi \pi $.\footnote{%
In the vacuum the process $\omega \to \pi ^{+}\pi ^{-}$ occurs with a 2\%
branching ratio. This is due to small isospin breaking and the resulting $%
\rho -\omega $ mixing. This is not the process we are concerned in this
paper, hence we work in the strict isospin limit.} This process has been
recently analyzed by Wolf, Friman, and Soyeur \cite{wolf}, where the $\omega
-\sigma $ mixing mechanism and the subsequent decay of the $\sigma $ into
two pions leads to a huge partial decay width for $\omega \to \pi \pi $, of
the order of a few hundred MeV at typical densities in relativistic
heavy-ion collisions, and for $\omega $ moving with the momentum of the
order of 500MeV with respect to nuclear matter. We carefully reanalyze the
calculation of Ref. \cite{wolf}. Firstly, we recognize that at the same
level of calculation (low-density expansion) there are additional diagrams
in nuclear medium, leading to $\omega \to \pi \pi $ without an intermediate $%
\sigma $ state (the first two diagrams in Fig. 2). The inclusion of these
diagrams provides cancellations with the diagram with the intermediate $%
\sigma $ state (third diagram in Fig. 2), which are strong when the mass of
the $\sigma $ is large. This can lead to a reduction of the in-medium
partial width for $\omega \to \pi \pi $ down to a level of a few tens of
MeV, which is much smaller than the value or Ref. \cite{wolf}, but still
large enough to be relevant among other processes contributing to the
in-medium $\omega $ width. More importantly, the inclusion of all diagrams
\begin{figure}[t]
\xslide{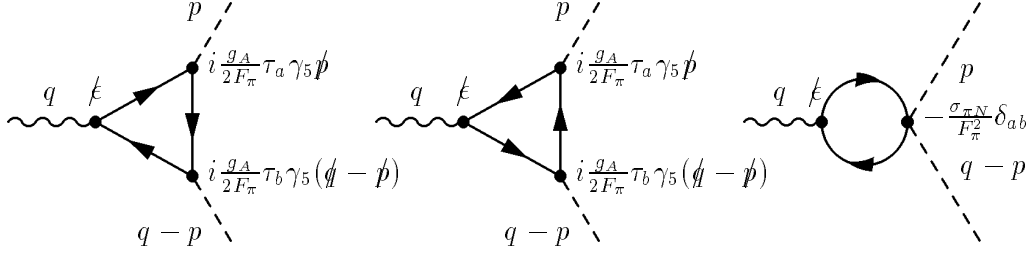}{38mm}{80}{395}{503}{507}
\label{pv}
\caption{One-nucleon-loop diagrams contributing to the $\omega \to \pi \pi$
amplitude in nuclear medium. Pseudovector coupling is used. The incoming
$\omega$ has momentum $q$ and polarization $\epsilon$. The outgoing pions
have momenta $p$ and $q-p$, and isospin $a$ and $b$, respectively.
Pseudovector coupling is used in the first two diagrams. The third diagram
contains the S-wave $\pi$-$N$ coupling, proportional to the sigma term,
$\sigma_{\pi N}$. The in-medium nucleon propagator is given in
Eq.~(\ref{Nprop}).}
\end{figure}
\begin{figure}[b]
\xslide{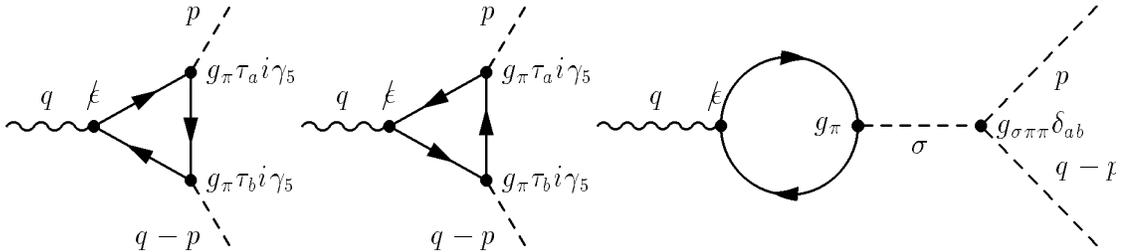}{38mm}{80}{395}{530}{507}
\label{ps}
\caption{Same as Fig. \ref{pv} evaluated in the linear $\sigma$-model. In
this case pseudoscalar $\pi$-$N$ coupling is used, and the third
diagram involves the intermediate $\sigma$ meson
propagator, $1/(m_\sigma^2-m_\omega^2)$. The
$\pi$-$N$ and $\sigma$-$N$ coupling constants are equal and
denoted by $g_\pi$. The $g_{\sigma\pi\pi}$ coupling constant is
equal to $(m_\sigma^2-m_\pi^2)/F_\pi$.}
\end{figure}
in Fig. 2 is necessary to satisfy the constraints of chiral symmetry in
formal limiting cases (small external 4-momenta). Also, we take the effort
to analyze separately the longitudinally and transversely polarized $\omega $%
, and show that the partial width for the former is a few times larger than
for the latter. We work for simplicity in the low-density expansion, which
allows us to develop simple and instructive formulas for the amplitude and
the decay width in the case when the 3-momentum of the $\omega $ meson is
small, and close to the chiral limit.

In hadronic calculations involving pions one has to choose whether to use
pseudovector or pseudoscalar $\pi N$ coupling. In the latter case one has
also to include the scalar-isoscalar $\sigma $ meson. In case of low-energy
physical processes, where all external four-momenta are small, the two
methods are equivalent and it is a matter of convenience which coupling to
use\cite{adler,alfaro,birse}. In the present case the four-momentum of the
on-shell $\omega $ meson is not small, therefore the two methods do not
yield the same result. Only in a formal limit of very large $\sigma $ mass
(or, equivalently, low $\omega$ mass) do the results overlap. We will
discuss this point very carefully throughout this Letter. Our calculation of
$\omega \to \pi \pi $ in nuclear matter includes the mechanism of absorption
of the $\omega $ meson by a nucleon from the Fermi sea, and emission of two
pions by this nucleon. These processes are depicted in Fig. 1 and Fig. 2,
respectively, for the case of pseudovector and pseudoscalar $\pi N$
coupling. The $\omega N$ coupling constant can be estimated from the vector
dominance model. We use $g_\omega =10$. The solid line in Figs. 1 and 2
denotes the in-medium nucleon propagator, which can be conveniently
decomposed in the free and density parts
\begin{equation}
G(k)\equiv G_F(k)+G_D(k)=(\FMSlash{k}+M)[\frac 1{k^2-M^2+i\varepsilon }+%
\frac{i\pi }{E_k}\delta (k_0-E_k)\theta (k_F-|\vec k|)].  \label{Nprop}
\end{equation}
The first two diagrams in Figs. 1 and 2 involve triangles with nucleon
lines. The only non-vanishing contributions involve one $G_D$ propagator and
two $G_F$ propagators. The third diagrams involve the bubbles with two
nucleon propagators, and are non-zero only when one propagator is $G_D$, and
the other one $G_F$. The third diagram in Fig. 1 contains the $\pi \pi NN$
vertex, which involves the pion-nucleon $\sigma $ term, $\sigma _{\pi
N}\simeq 45\mathrm{MeV}$. The process included in the third diagram in Fig.
2 involves the intermediate $\sigma $-meson propagator, which in the $\sigma
$-model takes the simple form $1/(q^2-m_\sigma ^2)$.

The amplitude, evaluated according to diagrams in Fig. 1 or Fig. 2 can be
decomposed in the following Lorentz-invariant way
\begin{eqnarray}
\mathcal{M}=\epsilon ^\mu (Ap_\mu +Bu_\mu +Cq_\mu ),  \label{decomp}
\end{eqnarray}
where $p$ is the momentum of one of the pions, $q$ is the momentum of the $%
\omega $ meson, $u$ is the four-velocity of nuclear matter, and $\epsilon$
specifies the polarization of $\omega$. For the case of the pseudovector
coupling (Fig.~1) the explicit calculation (in the rest frame of the nuclear
matter) yields:
\begin{eqnarray}
A^{\mathrm{PV}} &=&g_\omega \frac{g_A^2}{F_\pi ^{2_{}}}\frac{4(m_\omega
^2+2m_\pi ^2)\cos \gamma \ |\vec q|}{m_\omega ^2(m_\omega ^2-4M^2)}\rho
_B+...\hspace{0.5cm},  \label{apv} \\
B^{\mathrm{PV}} &=&g_\omega \left[ \frac{g_A^2}{F_\pi ^{2_{}}}\frac{%
2m_\omega ^2}{m_\omega ^2-4M^2}-\frac{\sigma _{\pi N}}{F_\pi ^{2_{}}}\frac{8M%
}{m_\omega ^2-4M^2}\right] \rho _B+...\hspace{0.5cm},  \label{bpv} \\
C^{\mathrm{PV}} &=&g_\omega \left[ - \frac{g_A^2}{F_\pi ^{2_{}}}\frac{%
2(m_\omega ^3+(m_\omega ^2+2m_\pi ^2)\cos \gamma \ |\vec q|)}{m_\omega
^2(m_\omega ^2-4M^2)} + \frac{\sigma _{\pi N}}{F_\pi ^{2_{}}}\frac{8M}{%
m_\omega (m_\omega ^2-4M^2)}\right] \rho _B+...\hspace{0.5cm}.  \label{cpv}
\end{eqnarray}
where $\gamma $ is the angle between $\vec q$ and $\vec p$, and $...$ mean
higher-order terms in the baryon density $\rho _B$, momentum $|\vec q|$, and
the chiral parameter $m_\pi $. We chose to present the results in the lowest
expansion in these three parameters due to simplicity. We should also
comment here that in fact it is not justified to keep terms in the density
expansion which are of order $\rho _B^2$ or higher, since these result from
diagrams not included in the calculation.\footnote{%
The next terms in the density expansion are the Fermi-motion correction,
which start at $\rho _B^{4/3}$. These could be included, but technically it
is rather involved for the triangle diagrams in Figs. 1 or 2. Hopefully,
these are not too large, as was the case \emph{e.g. } in model calculations
of the density-dependence of the quark condensate \cite{birse}.}

For the case of the pseudoscalar coupling (Fig. 2) we get analogously
\begin{eqnarray}
A^{\mathrm{PS}} &=&g_\omega \frac{g_\pi ^2}{M^2}\frac{4(m_\omega ^2+2m_\pi
^2)\cos \gamma \ |\vec q|}{m_\omega ^2(m_\omega ^2-4M^2)}\rho _B+...%
\hspace{0.5cm},  \label{aps} \\
B^{\mathrm{PS}} &=&g_\omega \frac{g_\pi ^2}{M^2}\left[ 2+\frac{8M^2(m_\sigma
^2-m_\pi ^2)}{(m_\omega ^2-4M^2)(m_\sigma ^2-m_\omega ^2)}\right] \rho _B+...%
\hspace{0.5cm},  \label{bps} \\
C^{\mathrm{PS}} &=&g_\omega \frac{g_\pi ^2}{M^2}\left[ -\frac{2(m_\omega
^3+(m_\omega ^2+2m_\pi ^2)\cos \gamma \ |\vec q|-4M^2m_\omega )}{m_\omega
^2(m_\omega ^2-4M^2)}\right.   \label{cps} \\
&&\left. -\frac{8M^2(m_\sigma ^2-m_\pi ^2)}{m_\omega (m_\omega
^2-4M^2)(m_\sigma ^2-m_\omega ^2)}\right] \rho _B+...\hspace{0.5cm}.
\nonumber
\end{eqnarray}
Eqs. (\ref{apv}-\ref{cpv}) and Eqs. (\ref{aps}-\ref{cps}) are equivalent if
the following conditions are satisfied: (i) the Goldberger-Treiman relation
holds, \emph{i.e.} $g_AM=g_\pi F_\pi $, (ii) the pion-nucleon sigma term
satisfies the $\sigma $-model relation $\sigma _{\pi N}=g_A^2M\ m_\pi
^2/m_\sigma ^2$, and (iii) ${m_\sigma ^2}/{m_\omega ^2}\rightarrow \infty $.
Condition (i) follows from chiral symmetry and is satisfied well with
experimental numbers. Condition (ii) and (iii) are formal. When imposed,
they lead the strong cancellations between the ``triangle'' and
``bubble'' diagrams, which can now be explicitly observed in the form of the
coefficients $B$ and $C$ in Eqs. (\ref{bps},\ref{cps}). In realistic cases,
however, the third condition is not well satisfied and the cancellations are
weaker. In the extreme situation, for $m_\sigma $ approaching $m_\omega $,
the $\omega $ decay width strongly increases and the $\sigma $ diagram
dominates. However, in this case the approach should include the finite
width of $\sigma $, which eliminates the divergence \cite{wolf}.

The expression for the decay width reads \cite{bjorken}
\begin{equation}
\Gamma _{\omega \rightarrow \pi \pi }=\frac 123\frac 1{n_s}\sum_s\frac
1{2q_0}\int \frac{d^3p}{(2\pi )^32p_0}\int \frac{d^3p^{\prime }}{(2\pi
)^32p_0^{\prime }}|\mathcal{M}|^2(2\pi )^4\delta ^{(4)}(q-p-p^{\prime }),
\label{widthg}
\end{equation}
where the factor $\frac 12$ is the symmetry factor when the decay proceed
into two neutral pions, the factor of $3$ accounts for the isospin
degeneracy of the final pion states (\emph{i.e.} neutral and charged pions),
$n_s$ is the number of spin states of the $\omega $ meson, and $\sum_s$
denotes the sum over these spin states, and $q$ , $p$ and $p^{\prime }=q-p$
are the four-momenta of the $\omega $ meson, and the two pions, respectively
(cf. Figs. 1 and 2). The phase-space integral is performed in the rest frame
of the nuclear medium, and we obtain explicitly
\begin{equation}
\Gamma _{\omega \rightarrow \pi \pi }=\frac 123\frac 1{n_s}\sum_s\frac
1{2q_0}\int \sin \gamma \frac{\vec p^2}{8\pi p_0(q_0-p_0)|a|}|\mathcal{M}%
|^2d\gamma ,  \label{width}
\end{equation}
where the kinematics enforces
\begin{eqnarray}
|\vec p| &=&\frac{m_\omega ^2|\vec q|\cos \gamma +q_0\sqrt{m_\omega
^4-4m_\pi ^2(m_\omega ^2+\vec q^2\sin ^2\gamma )}}{2(m_\omega ^2+\vec
q^2\sin ^2\gamma )},\hspace{0.5cm}q_0=\sqrt{m_\rho ^2+\vec q^2},  \nonumber
\\
\qquad p_0 &=&\sqrt{m_\pi ^2+\vec p^2},\hspace{0.5cm}a=\left. \frac{d(q_0-%
\sqrt{m_\pi ^2+r^2}-\sqrt{m_\pi ^2+r^2-2r|\vec q|\cos \gamma +\vec q^2})}{dr}%
\right| _{r=|\vec p|}\!\!\!\!\!\!.
\end{eqnarray}
The presence of the medium results in the splitting of transversely and
longitudinally polarized $\omega $ states. Transversely polarized $\omega $
has two helicity states ($n_s=2$), with projection $s=\pm 1$ on the
direction of $\vec q$, and the longitudinally polarized $\omega $ has one
helicity state ($n_s=1$), with the corresponding projection $s=0$. An
explicit calculation yields
\begin{eqnarray}
\sum_{s=\pm 1}\varepsilon _{(s)}^\mu \varepsilon _{(s)}^\nu  &=&g^{\mu \nu
}-u^\mu u^\nu -\frac{(q^\mu -q\cdot u\ u^\mu )(q^\nu -q\cdot u\ u^\nu )}{%
q\cdot q-(q\cdot u)^2}\equiv T^{\mu \nu },  \nonumber \\
\varepsilon _{(s=0)}^\mu \varepsilon _{(s=0)}^\nu  &=&-\frac{q^\mu q^\nu }{%
q\cdot q}+u^\mu u^\nu +\frac{(q^\mu -q\cdot u\ u^\mu )(q^\nu -q\cdot u\
u^\nu )}{q\cdot q-(q\cdot u)^2}\equiv L^{\mu \nu }.  \label{tensors}
\end{eqnarray}
Note that by summing over all polarization we recover
\mbox{$\sum_{s=0,\pm 1}\varepsilon _{(s)}^\mu \varepsilon _{(s)}^\nu =g^{\mu
\nu }-\frac{q^\mu q^\nu }{q\cdot q}$}. The tensors $T^{\mu \nu }$ and $%
L^{\mu \nu }$ are projection tensors, \emph{i.e. }, $T^{\mu \nu }T_\nu
^{\cdot \alpha }=T^{\mu \alpha }$, $L^{\mu \nu }L_\nu ^{\cdot \alpha
}=L^{\mu \alpha }$, and $T^{\mu \nu }L_\nu ^{\cdot \alpha }=0$. Furthermore,
$T^{\mu \nu }q_\nu =0$ and $L^{\mu \nu }q_\nu =0$, which reflects current
conservation, as well as $T^{\mu \nu }u_\nu =0$. Using relations (\ref
{decomp},\ref{tensors}) in Eq. (\ref{width}) we find the following
expressions for the decay widths of transversely and longitudinally
polarized $\omega $ meson into two pions:
\begin{equation}
\Gamma _{\omega \rightarrow \pi \pi }^T=A^2p_\mu T^{\mu \nu }p_\nu
,\;\;\;\;\Gamma _{\omega \rightarrow \pi \pi }^L=(Ap_\mu +Bu_\mu )L^{\mu \nu
}(Ap_\nu +Bu_\nu ).  \label{widthL}
\end{equation}
We note that the coefficient $C$ does not enter these formulas. For the case
of pseudovector coupling we obtain:
\begin{eqnarray}
\Gamma _{\omega \rightarrow \pi \pi }^{T,PV} &=&\frac{g_\omega ^2}{40\pi }%
\left( \frac{g_A}{F_\pi }\right) ^4\frac{m_\omega ^2-2m_\pi ^2}{m_\omega
(m_\omega ^2-4M^2)^2}\vec q^2\rho _B^2+...\hspace{0.5cm},  \label{gtpv} \\
\Gamma _{\omega \rightarrow \pi \pi }^{L,PV} &=&\frac{g_\omega ^2}{5\pi }%
\left( \frac{g_A}{F_\pi }\right) ^4\frac{m_\omega ^2-2m_\pi
^2-10Mg_A^{-2}\sigma _{\pi N}}{m_\omega (m_\omega ^2-4M^2)^2}\vec q^2\rho
_B^2+...\hspace{0.5cm},  \label{glpv}
\end{eqnarray}
where $...$ denote higher-order terms in the expansion in $\rho _B$, $\vec q$%
, and $m_\pi $. We note that if the term with $\sigma _{\pi N}$ (which is of
the order $m_\pi ^2$) were neglected, then the width of the longitudinal
mode, (\ref{glpv}), would be 8 times larger than the width of the transverse
mode, (\ref{gtpv}). However, numerically we find $-10Mg_A^{-2}\sigma _{\pi
N}/(m_\omega ^2-2m_\pi ^2)\simeq -0.3$, which is large, and our expansion
gives $\Gamma _{\omega \rightarrow \pi \pi }^{L,PV}/$ $\Gamma _{\omega
\rightarrow \pi \pi }^{T,PV}\simeq 5$. This indicates that with the physical
values of the parameters we are not very close to the chiral limit.

For the case of pseudoscalar coupling we find
\begin{eqnarray}
\Gamma _{\omega \rightarrow \pi \pi }^{T,PS} &=&\frac{g_\omega ^2}{40\pi }%
\left( \frac{g_\pi }M\right) ^4\frac{m_\omega ^2-2m_\pi ^2}{m_\omega
(m_\omega ^2-4M^2)^2}q^2\rho _B^2+...\hspace{0.5cm},  \label{gtps} \\
\Gamma _{\omega \rightarrow \pi \pi }^{L,PS} &=&\frac{g_\omega ^2}{5\pi }%
\left( \frac{g_\pi }M\right) ^4\left\{ \frac{m_\omega ^2-2m_\pi
^2-10M^2m_\pi ^2/m_\sigma ^2}{m_\omega (m_\omega ^2-4M^2)^2}+\right.
\label{glps} \\
&&~\hspace{-11mm}\left. \frac{10M^2[(3M^2-m_\omega ^2+m_\sigma ^2)m_\omega
^2+(m_\omega ^4/m_\sigma ^2-12M^2+m_\omega ^2-2m_\sigma ^2)m_\pi ^2]}{%
m_\omega (m_\omega ^2-4M^2)^2(m_\omega ^2-m_\sigma ^2)^2}\right\} q^2\rho
_B^2+...\hspace{0.5cm}.  \nonumber
\end{eqnarray}
We note that Eqs. (\ref{gtps}) and (\ref{gtpv}) are equal. The expression
for the width of the longitudinal mode, Eq. (\ref{glps}), is written in such
a way that the first term in the curly brackets reproduces the
pseudovector-coupling result, (\ref{glpv}), when $\sigma _{\pi N}=g_A^2M\
m_\pi ^2/m_\sigma ^2$. The second term in the curly brackets in (\ref{glps})
comes entirely from the $\sigma $-diagram in Fig. 2. Formally, it vanishes
when $m_\sigma ^2/m_\omega ^2\rightarrow \infty $. It\ becomes very large
when the positions of the $\sigma $ and $\omega $ resonances are close to
each other, even when we supply $\sigma $ with finite width, as in Ref. \cite
{wolf}.

\begin{figure}[tbp]
\vspace{0mm} \epsfxsize = 9 cm \centerline{\epsfbox{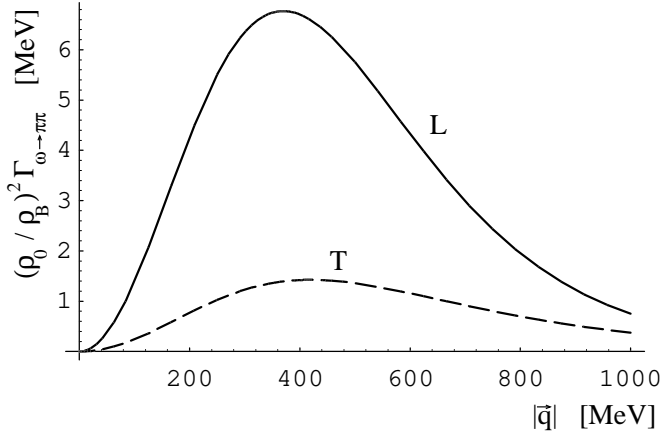}} \vspace{0mm}
\label{diag}
\caption{The quantity $(\rho_0/\rho_B)^2 \Gamma_{\omega \to \pi\pi}$ plotted
as a function of the 3-momentum of the $\omega $ meson, $|\vec q|$. The
solid (dashed) line correspond to the longitudinal (transverse) mode.}
\end{figure}

In Fig. 3 we present numerical results for the pseudovector-coupling case
for \emph{finite} $q$ and $m_\pi $, but still in the low-density
approximation. We plot the quantities $(\rho _0/\rho _B)^2\Gamma _{\omega
\rightarrow \pi \pi }^{T,PV}$ and $(\rho _0/\rho _B)^2\Gamma _{\omega
\rightarrow \pi \pi }^{L,PV}$, where $\rho_0$ is the nuclear saturation
density, as functions of the momentum $|\vec q|$. This way we get rid of the
density dependence. Also, in order to obtain more reasonable numerical
estimates, we depart in this calculation from the strict low-density limit
by reducing the value of the nucleon mass to 70\% of its vacuum value, which
is a typical in-medium number (\emph{i.e.} $M^{*}/M=0.7$). This is a very
important effect, since expressions (\ref{gtpv}-\ref{glps}) scale
approximately as $M^{-8}$. The density dependence on other quantities is
less crucial, so we do not take it into account. The plot shows that the
width of the longitudinal mode is a few times larger than the width of the
transverse mode for all values of $|\vec q|$. We note strong dependence on $%
|\vec q|$, with maxima around 400MeV. The values around the maxima for $\rho
_B=2\rho _0 $ give $\Gamma _{\omega \rightarrow \pi \pi }^{T,PV}\simeq 5%
\mathrm{MeV}$ and $\Gamma _{\omega \rightarrow \pi \pi }^{L,PV}\simeq 25%
\mathrm{MeV}$. The latter value is substantial, showing that the $\omega
\rightarrow \pi \pi $ process should be included (among other processes) in
the analysis of the in-medium modification of the $\omega $ meson.

In conclusion, we would like to remark on the possibly large role of the
final-state interactions. In fact, the $\sigma $-diagram of Fig.~2 may be
viewed as an example of a process where the pions form a resonance in the
final state. This can lead to a large enhancement of the amplitude, as was
the case in Ref. \cite{wolf}, and which is explicit in our expressions when $%
m_\sigma $ is close to $m_\omega $. However, final-state interactions should
be included in all diagrams of Fig.~2 (or Fig.~1), hence cancellations are
expected to  occur. This interesting extension of the present analysis is
left for a future work.


\end{document}